\documentclass[
reprint,
superscriptaddress,
twocolumn,
aip,
cha,
10pt,
]{revtex4-1}

\usepackage[utf8]{inputenc}
\usepackage{graphicx}
\usepackage{braket}
\usepackage{amssymb}
\usepackage{amsmath} 
\usepackage{lipsum}
\usepackage{siunitx}
\usepackage{empheq}

\pdfoutput=1

\begin{document}

\title{Optical Non-Linearities in Plasmon--Exciton Core--Shell Systems: The Role of Heat}

\author{Felix Stete}
\affiliation{Institut für Physik \& Astronomie, Universität Potsdam, Karl-Liebknecht-Str. 24-25, 14476 Potsdam, Germany}
\affiliation{School of Analytical Sciences Adlershof (SALSA), Humboldt-Universität zu Berlin, Unter den Linden 6, 10999 Berlin, Germany}
\author{Matias Bargheer}
\affiliation{Institut für Physik \& Astronomie, Universität Potsdam, Karl-Liebknecht-Str. 24-25, 14476 Potsdam, Germany}
\affiliation{Helmholtz Zentrum Berlin, Albert-Einstein-Str. 15, 12489 Berlin, Germany}
\author{Wouter Koopman}
\affiliation{Institut für Physik \& Astronomie, Universität Potsdam, Karl-Liebknecht-Str. 24-25, 14476 Potsdam, Germany}

\begin{abstract}
\normalsize 
\textbf{Abstract:}
Strong coupling between plasmons and excitons gives rise to new hybrid polariton states with various fields of potential applications. Despite a plethora of research on plasmon--exciton systems, their transient behaviour is not yet fully understood. Besides Rabi oscillations in the first femtoseconds after an optical excitation, coupled systems show interesting non-linear features on the picosecond time scale. Here, we conclusively show that the source of these features is heat that is generated inside the particles. Until now, this hypothesis was only based on phenomenological arguments. We investigate the role of heat by recording transient spectra of plasmon--exciton core--shell nanoparticles with excitation off the polariton resonance. We present analytical simulations that precisely recreate the measurements solely by assuming an initial temperature rise of the electron gas inside the particles. The simulations combine established strategies for describing uncoupled plasmonic particles with a recently published model for static spectra. The simulations are consistent for various excitation powers confirming that indeed heating of the particles is the root of the changes in the transient signals.
\end{abstract}

\maketitle


\noindent 
The small mode volume of localized surface plasmons on gold nanoparticles greatly enhances external electric fields. This enhancement enables high coupling strengths with excitonic emitters on the particle surface. In the so-called strong coupling regime, new hybrid resonances emerge, an upper and a lower polariton mode. These polaritons possess both plasmonic and excitonic characteristics, a feature that is of interest for applications like quantum networks, \cite{Imamoglu.1999} parametric optical signal amplification, \cite{Saba.2001} manipulation of chemical reactions, \cite{Hutchison.2012} thresholdless lasing \cite{Torma.2014} and many more.

Strong coupling between excitonic emitters and metal nanoparticles has been realized for various particle types. Examples are silver rods, \cite{Zheng.2017} silver triangles, \cite{Zengin.2015} silver shells, \cite{Zhou.2016} gold rods, \cite{Melnikau.2016} aluminum discs. \cite{Eizner.2015} When excited, the energy oscillates between purely plasmonic and purely excitonic modes at the Rabi frequency. For metal nanoparticles, this frequency is on the order of a few fs making a direct observation difficult. Additionally, the short life time of plasmons on the order of 10\,fs hinders a long coherent oscillation between excitonic and plasmonic parts. The strongly damped Rabi oscillations have been observed for surface plasmon polariton systems \cite{Vasa.2013} and only very recently, coherent coherent dynamics have been observed in colloidal systems. \cite{Peruffo.2023}
 
Also on longer time scales after an initial excitation, strongly coupled plasmon--exciton nanoparticles show interesting non-linear features that can be of great importance for future applications. Several studies have discussed the transient behaviour of coupled plasmon--exciton systems on a time for scale beyond the polariton life time.\cite{Hao.2011,Simon.2016,Eizner.2017,Finkelstein-Shapiro.2021} 
The main source for changes in the transient spectra at the polariton resonances was argued to be heat in the metal cores. But the arguments were rather phenomenological than rigid. In particular, the modes were modelled as simple oscillators. However, for a detailed investigation of the role of heat in the system, the material properties of the particle need to be taken into account.

For bare nanoparticles, such a complete approach has successfully been introduced to explain changes of optical properties upon heating or the transient behaviour after an initial excitation.\cite{Voisin.2001, Hodak.1998, Park.2007} The main idea is to model the metal's permittivity depending on the electron and lattice temperature in the particle since for bare particles, heat is the source of changes in the transient spectra. This model has not yet been applied to the situation of plasmon--exciton particles to precisely discuss the role of heat in the metal.

We want to close this gap with this study by presenting simulations of the transient spectra of coupled plasmon--exciton nanoparticles which are exited off the polariton resonance. We combine, on the one hand, the well-established method of reproducing the transient behaviour of gold nanoparticles via Rosei's model \cite{Stoll.2014, Park.2007} and, on the other hand, a recently presented expansion of the Mie--Gans model to describe the static extinction of coupled core--shell particles taking into account the power broadening of the emitters induced by the plasmonic cavity.\cite{Stete.2020} The simulations essentially model how the increased electron temperature changes the occupied electronic states according to the Fermi-distribution and the lattice temperature adds a correction to the free electron damping. We calibrate the simulations by comparison to static spectra of the bare gold particles and core--shell particles. Finally, we model the electron and phonon temperatures in a two-temperature model and compare the simulated spectra to the results of pump--probe measurements of coupled plasmon--exciton core--shell nanorods. Our model system consists of a gold core and shell of the cyanine dye TDBC. The particles show strong coupling or at least verge on the strong coupling regime. \cite{Stete.2017,Stete.2018} The comparison between simulations and measurements for various pump powers demonstrates good agreement supporting previous studies which had to argue phenomenologically. \cite{Simon.2016, Hao.2011}

\section*{Experimental Methods} \vspace{-3.5mm}

\begin{figure}
\centering
\includegraphics[width=64mm]{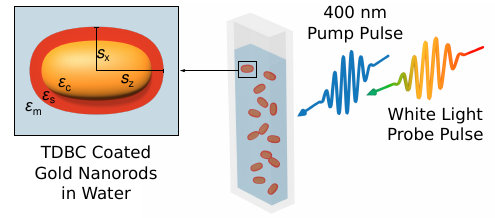}
\caption{Sketch of the experimental setting: TDBC coated gold nanorods were dissolved in water and excited by a $\SI{400}{nm}$ pump pulse. A white light probe pulse detected the change in transmission after excitation. A magnified particle is shown on the left with the parameters used later in the simulations.}
\label{fig:Sketch}
\end{figure}

\noindent 
We measured the change in transmission of white light laser pulses through a solution of dye coated gold nanorods for various delay times after excitation with a fs-pump pulse at 400\,nm (Figure \ref{fig:Sketch}).

\textbf{TDBC-coated gold nanorods}: 
The cyanine dye 5,\,5',\,6,\,6'-tetrachloro-1-1'-diethyl-3,\,3'-di\,(4-sulfobutyl)-benzimidazolocarbocyanine (TDBC) was obtained from \textit{FEW chemicals} and used as received. It was dissolved in an aqueous NaOH solution ($c_{\textrm{NaOH}}=\SI{e-5}{mol/l}$) with a concentration of $c_{\textrm{TDBC}}=\SI{1}{mmol/l}$, stirred for $\SI{5}{min}$ and placed in an ultrasonic bath for $\SI{15}{min}$. 
Citrate stabilized gold nanorods with a nominal transverse diameter of 25\,nm and a longitudinal plasmon resonance at 600\,nm were purchased from \textit{Nanopartz}. Contrary to CTAB ligands, conventionally used for stabilizing gold nanorods, citrate ligands can easily be exchanged by TDBC. 
Particle and dye solution were mixed in a ratio of 1:1, placed in an ultrasonic bath for $\SI{15}{min}$ and left undisturbed for $\SI{48}{h}$. After this resting time, the unadsorbed TDBC was removed by centrifuging twice for $\SI{30}{min}$ at $\SI{3000}{rpm}$ and refilling with purified water. During the measurements, the particle solution was filled into cuvettes with a path length of $\SI{1}{mm}$. 

\textbf{Static measurements}: Static transmission spectra were recorded with a \textit{Cary 5e} spectrometer.

\textbf{Pump--probe measurements}: Laser pulses with a pulse length of $\SI{140}{fs}$ at a central wavelength of $\SI{795}{nm}$ and a repetition rate of $\SI{5}{kHz}$ were generated in a Ti:sapphire laser system (\textit{MaiTai/Spitfire Pro} by \textit{Spectra-Physics}). A part of the light was frequency doubled and focussed on the particle solution as pump beam with a focal spot size of approximately $\SI{300}{\micro m}$. The power was varied between $\SI{1}{mW}$ and $\SI{4.5}{mW}$. A small part of the output of the laser system was used to generate super continuum (white light) pulses in a $\SI{1}{mm}$ thick sapphire plate. These white light pulses were used to probe the transmission through the sample with an arbitrary delay to the pump excitation. After correction for the chirp of the white light as measured by polarisation-gated frequency-resolved optical gating (PG-FROG), the white light conserves the time resolution of the pump pulse. The time delay $t$ between pump and probe could be controlled with a delay stage. The pump beam was chopped at rate of $\SI{125}{Hz}$ to measure the relative change in transmission between the pumped ($T_0 + \Delta T$) and the unpumped ($T_0$) sample. The change in transmission is directly connected to the change in extinction $\Delta \sigma_\textrm{ext}$ of the particles via $\Delta T(t) \approx n L \Delta \sigma_\textrm{ext}(t)$ with $n$ representing the particle density and $L$ the sample thickness. \cite{Stoll.2014} 

\section*{Simulations} \vspace{-3.5mm}

\noindent 
This study discusses simulations of transient spectra of core--shell nanoparticles based on the heat-induced change of the core's permittivity $\epsilon_\textrm{c}$. For simplicity, we assume here that the excitonic shell permittivity $\epsilon_\textrm{c}$ is not affected by the excitation. 

The simulation procedure was perfomed as follows: In a first step, we simulated the static spectra at room temperature by combining the theoretical description of the gold permittivity at room temperature by Rosei \cite{Rosei.1974,Rosei.1973,Guerrisi.1975,Winsemius.1975} with the Mie--Gans model for core--shell nanospheroids. \cite{Stete.2020} Transient spectra were then obtained by modeling the effect of the transient temperature rise on the Rosei-permittivity. The single steps on this route are described in detail below.

\textbf{Static spectra}: We simulated the static spectrum of our particles based on the Mie-Gans model. \cite{Stete.2020} The extinction cross section $\sigma_\textrm{ext}$ of a nanoparticle can be obtained by calculating its polarisability $\alpha$. In a dipolar approximation, the polarisability $\alpha_{0,j}$ along the long ($j=z$) or the short ($j=x,y$) axis of a spheroid is given by \cite{Bohren.2008,Zengin.2013}
\begin{equation}
\begin{split}
\alpha_{0,j} =  V \frac{(\epsilon_\textrm{s}-\epsilon_\textrm{m}) \epsilon_\textrm{a} + g \epsilon_\textrm{s}(\epsilon_\textrm{c}-\epsilon_\textrm{s})}
{(\epsilon_\textrm{m}+L^{(2)}_j (\epsilon_\textrm{s}-\epsilon_\textrm{m}))\epsilon_\textrm{a} + g L^{(2)}_j \epsilon_\textrm{s}(\epsilon_\textrm{c}-\epsilon_\textrm{s})}\:.
\end{split}
\label{eq:Polarisability}
\end{equation}
Here, $\epsilon_\textrm{a} = \epsilon_\textrm{s} + (\epsilon_\textrm{c} - \epsilon_\textrm{s})(L^{(1)}_j - g L^{(2)}_j)$ with  $\epsilon_\textrm{c}$, $\epsilon_\textrm{s}$ and $\epsilon_\textrm{m}$ representing the permittivity of the core, the shell or the surrounding medium, respectively (Figure \ref{fig:Sketch}). $g$ describes the volume fraction of the core and $L^{(1,2)}_j$ the geometrical factors of the inner (1) and outer (2) spheroid.

To account for the finite particle size, we applied the \textit{modified long wavelength approximation} (MLWA) as \cite{Kelly.2003}
\begin{equation}
\alpha_j = \alpha_{0,j} \left( 1 - i \frac{\alpha_{0,j}}{6 \pi \epsilon_0} k^3 - \frac{\alpha_{0,j}}{4 \pi \epsilon_0} \frac{k^2}{s_j}
\right)^{-1}
\label{eq:MLWA}
\end{equation}
where $k$ is the wave vector in the medium and $s_j$ the respective semiaxis. Subsequently, an expression for the particle extinction was derived as
\begin{equation}
\sigma_\textrm{ext,j} = \frac{k^4}{6 \pi \epsilon_0^2} |\alpha_j|^2 + \frac{k}{\epsilon_0} \text{Im}(\alpha_j)\:.
\label{eq:SigmaExt}
\end{equation}
For this study, the particles were dissolved in water and thus randomly oriented in reference to the light polarisation. This is taken into account by weighing the contribution of each main axis with a factor of $1/3$. 

Apparently, the extinction is directly related to the permittivities $\epsilon_i$ of the involved materials. In the following, we want to present the models for $\epsilon_\textrm{s}$ and $\epsilon_\textrm{c}$.

For a correct simulation of the shell permittivity $\epsilon_\textrm{s}$, the power broadening of the molecular transition by the strong electric fields at the particle surface has to be taken into account. The permittivity of a two-level system in a cavity is given by \cite{Stete.2020} 
\begin{equation}
\epsilon_\textrm{s} = \epsilon_\infty + \frac{f\omega_0}{2} \frac{\omega_0-\omega+i\frac{\gamma}{2}}{(\omega_0-\omega)^2+\frac{\gamma^2}{4}+\frac{\Omega_0^2}{2}}\:.
\end{equation}
Here, $f$, $\gamma$ and $\omega_0$ represent the system's oscillator strength, linewidth and resonance position, respectively while $\Omega_0$ denotes the vacuum Rabi frequency, quantifying the molecule-particle coupling.

A theoretical model for the temperature-dependence of the Au-permittivity was established by Rosei and colleagues \cite{Rosei.1974, Rosei.1973, Guerrisi.1975, Winsemius.1975} and has successfully been used to describe the transient behaviour of gold nanoparticles. \cite{Stoll.2014, Park.2007} It includes the contribution of the conduction electrons $\epsilon_\textrm{Dr}$ via the Drude model as well as the contribution of interband transitions $\epsilon_\textrm{IB}$ via the joint density of states. Combining the different contributions allows the calculation of the imaginary part $\text{Im}(\epsilon)$. The real part of the permittivity $\text{Re}(\epsilon)$ is then obtained via the Kramers-Kronig integral.

The optical response of the quasi-free electrons at the Fermi edge is described by the Drude contribution to the permittivity as
\begin{equation}
\epsilon_\textrm{Dr}(\omega) = 1 + \frac{\omega_\textrm{p}^2}{\omega^2 + i\gamma_\textrm{Dr} \omega}\:.
\end{equation}

\begin{figure}
    \centering
    \includegraphics{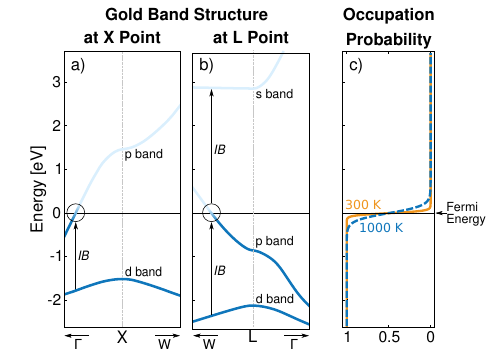}
    \caption{Band structure of gold at the X (a) and L (b) point in the Brillouin zone and the corresponding occupation probability given by the Fermi distribution $f(E,T)$ (c). Vertical arrows indicate the interband transitions (IB) with transition energies in the visible. The Fermi distribution is plotted for $T=\SI{300}{K}$ (solid orange line) and $T=\SI{1000}{K}$ (dashed blue line) to illustrate the smearing of the Fermi edge.}
    \label{fig:GoldBandStructure}
\end{figure}

To correctly describe the permittivity of gold in the visible region down to $\SI{400}{nm}$, three interband transitions need to be taken into account in addition to the Drude contribution: The transition from the d band to the p band close to the X point of the Brillouin zone (Figure \ref{fig:GoldBandStructure}a) and the transitions from the d band to the p band and from the p band to the s band in the vicinity of the L point of the Brillouin zone \cite{Stoll.2014} (Figure \ref{fig:GoldBandStructure}b). The basis for the contribution of the transition from band $i$ to band $j$ is the energy distribution of the joint density of states $D_{i\rightarrow j}(E,\hbar \omega)$.\cite{Rosei.1974} It describes the density of transitions with energy $\hbar \omega$ and initial energy $E$. To find the probability of a transition this term needs to weighted with the probability that the initial state is occupied while the final state is not. This occupation probability is determined by the Fermi distribution $f(E,T)$ whose edge at the Fermi energy changes with temperature (Figure \ref{fig:GoldBandStructure}c). The transition probability from band $i$ to band $j$ with energy $\hbar \omega$ is then \cite{Stoll.2014}
\begin{equation}
\begin{split}
J_{i\rightarrow j}(\hbar \omega) = & \int_{E_\textrm{min}}^{E_\textrm{max}} D_{i\rightarrow j}(E,\hbar \omega) \\
& \times  [f(E,T)(1-f(E+\hbar \omega,T))] dE  \:.
\end{split}
\label{eq:JDOS}
\end{equation}

Each interband transition is weighted with an oscillator strength $A_{i \rightarrow j}$ \cite{Rosei.1974} to model the imaginary part of the permittivity. Considering the contribution of the three relevant transitions, $\epsilon_\textrm{IB}$ then reads \cite{Stoll.2014}
\begin{equation}
\begin{split}
\text{Im}(\epsilon_\textrm{IB}(\omega)) = & \frac{4 \pi^2 q_\textrm{e}^2}{\epsilon_0 m_\textrm{e}^2 \omega^2} 
(A_{d \rightarrow p}^\textrm{X} J_{d\rightarrow p}^\textrm{X}(\omega) \\ 
& + A_{d \rightarrow p}^\textrm{L} J_{d\rightarrow p}^\textrm{L}(\omega) + 
 A_{p \rightarrow s}^\textrm{L} J_{p\rightarrow s}^\textrm{L}(\omega))\:.
\end{split}
\end{equation}

Combining Drude and interband contributions, we obtain an \textit{ab initio} expression for the imaginary part of the permittivity of gold. The real part is then directly retrieved via Kramers-Kronig integration. The necessary parameters for the model can be retrieved from the detailed calculated  band structure of gold. \cite{Stoll.2014} However, for simplicity and higher precision, they are usually found by fitting the permittivity model to the experimental data. For this study, we used the values provided by Ref\,\citenum{Olmon.2012}.

\textbf{Temperature dependent spectra}: The Rosei model allows simulating the temperature dependence of the permittivity. Heating causes a smearing of the Fermi edge (Figure \ref{fig:GoldBandStructure}c) and thus affects $J_{i\rightarrow j}(\hbar \omega)$ as calculated in eq \ref{eq:JDOS}. In other words, heating alters the occupation probability of the final or initial state and thus affects the different interband transitions. The Drude part is affected by heating via a change in the scattering rate $\gamma_\textrm{Dr}$. For weak and moderate excitation, the change in electron--electron scattering can be negelcted  and $\gamma_\textrm{Dr}$ is purely modified by a change in the electron--phonon scattering \cite{Rethfeld.2017} i.e. by the phonon temperature $T_\textrm{ph}$.

To simulate the effect of transient gold heating, we need to find expressions for the time dependent electron and phonon temperatures, $T_\textrm{e}(t)$ and $T_\textrm{ph}(t)$, which are incorporated into the Rosei model to determine the time dependent permittivity $\epsilon_\textrm{core}(t)$ of the gold. Subsequently, $\Delta \sigma_\textrm{ext}(t)$ is calculated via Equations \ref{eq:Polarisability}--\ref{eq:SigmaExt} and compared to the measured data. 

In metals, the energy of an incoming laser pulse is absorbed only by the electrons. They thermalize within a few tens of femtoseconds. Due to the fast thermalisation (in comparison to the temporal resolution of the set-up), we can assume here that directly after the excitation, all electrons are described by a Fermi distribution with temperature $T_\textrm{e,0}$. This temperature is derived from the absorbed pulse energy and the particle's absorption cross section. Since the interaction between the lattice and light can be neglected, the electrons and the lattice need to be described with two different temperatures which equilibrate via electron--phonon coupling in the first few picoseconds after excitation. Since both subsytems contribute to the permittivity (the electron temperature to the interband transitions, the phonon temperature to the Drude part), a two-temperature model is required to describe the transient behaviour of metal nanoparticles. The temperatures $T_i(t)$, where $i$ represents the electron (e) or lattice (l) system, are connected via the system of coupled differential equations \cite{Hodak.1998}
\begin{subequations}
\begin{empheq}{align}
       C_\textrm{e}(T_\textrm{e}) \frac{\partial T_\textrm{e}}{\partial t} &  = -G_\textrm{e--ph}(T_\textrm{e}-T_\textrm{l})
       \:, \\
        C_\textrm{l}(T_\textrm{l}) \frac{\partial T_\textrm{l}}{\partial t} &  = G_\textrm{e--ph}(T_\textrm{e}-T_\textrm{l})
        \:.
\end{empheq}
\label{eq:TwoTemperatureModel}
\end{subequations}
The two-temperature model is parametrized by the electron and lattice heat capacities $C_i(T_i)$, as well as the electron-phonon coupling constant $G_\textrm{e--ph}$. Coupling to the environment has been neglected in this expression since it occurs on longer times scales than the ones discussed here. The heat capacity of the gold lattice is $C_\textrm{l} = 3 k_\textrm{B} n$ with $k_\textrm{B}$ as Boltzmann constant and $n = \SI{5.9e-28}{m^{-3}}$. \cite{Park.2007} The electron heat capacity is temperature dependent via $C_\textrm{e} = \gamma_\textrm{e} T_\textrm{e}$ with a Sommerfeld constant of $\gamma_\textrm{e} = \SI{71.5}{Jm^{-3}K^{-2}}$. \cite{Park.2007} As electron phonon coupling, we use $G_{\textrm{e--ph}}=\SI{2.1e16}{W/m^3}$. \cite{Hohlfeld.2000}

The solution of these equations yields the two temperatures and consequently allows for a calculation of the permittivity at any time after excitation. The electron temperature determines the form of the Fermi distribution while the phonon temperature affects the free electron scattering rate $\gamma_\textrm{Dr}$ as \cite{Stoll.2014}
\begin{equation}
\frac{\Delta \gamma_\textrm{Dr}}{\gamma_{\textrm{Dr},0}} = \frac{\Delta T_\textrm{ph}}{T_0}  \:.
\end{equation}

In conclusion, this model allows the simulation of the permittivity of gold and consequently of the strongly coupled plasmon--exciton system after an initial excitation. In the following section, we will present the results of both measurement and simulation and their remarkable match.

\section*{Results and Discussion} \vspace{-3.5mm}

\begin{figure}[t]
    \centering
    \includegraphics{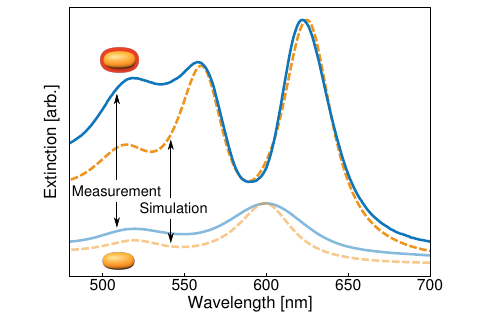}
    \caption{Measured and simulated static extinction spectra of bare and TDBC-coated gold nanorods. The solid blue lines represent the experimental data, the dashed orange lines represent the simulations. The two lower lines describe the bare particles whereas the two upper lines describe the coupled situation.}
    \label{fig:StaticSpectra}
\end{figure}

\noindent 
The static extinction spectrum of the dye coated nanorods shows three peaks (upper, dark blue line in Figure \ref{fig:StaticSpectra}). The transverse plasmon peak is located around $\SI{520}{nm}$ while around $\SI{570}{nm}$ and $\SI{630}{nm}$, the hybrid resonances of the coupled plasmon--exciton system become visible. For comparison, the extinction spectrum of bare gold nanorods (lower, bright blue line in Figure \ref{fig:StaticSpectra}) shows a longitudinal resonance around $\SI{600}{nm}$. The splitting in the plasmon--exciton spectrum originates from the strong coupling of this longitudinal plasmon resonance to the exciton resonance of TDBC around $\SI{600}{nm}$. The hybrid coupled excitations are known as the upper and lower polariton peak.

The static simulations are represented by the dashed orange lines in Figure \ref{fig:StaticSpectra}. The parameters used here are identical to those in Reference\,\citenum{Stete.2020}. However, for this study, the gold permittivity is now not directly taken from measured data \cite{Olmon.2012} but simulated via the Rosei model that fully reproduces this measured permittivity. The simulations of the extinction of the bare and coupled systems agree quite well with the data. Just the intensity of the transverse resonance is not fully recovered. This is a well known problem in simulations of nanorod spectra. \cite{Myroshnychenko.2008, Huanjun.2013}
In this case, nanospheres in the sample as well as particle clusters that can both be observed in SEM images seem to be the main reason for a higher peak in the measured data. In the region of interest i.e. the region of the coupled resonances, the model fits the data very well. We hence conclude that our model is suitable to further investigate the transient behavior of the coupled system.

\begin{figure}[t]
\centering
\includegraphics{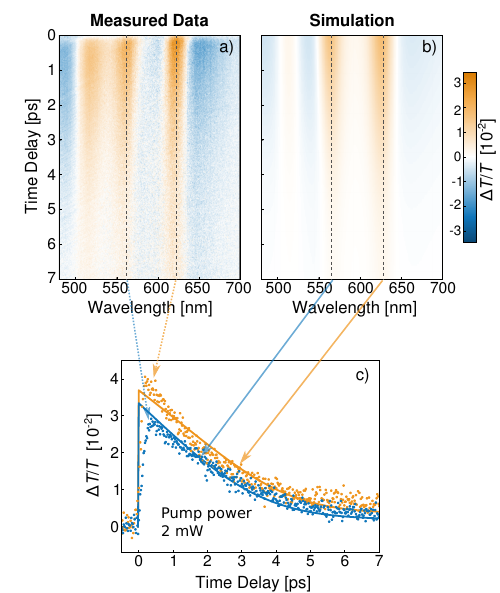}
\caption{Measured and simulated transient pump--probe spectra of gold--TDBC core--shell nanorods pumped with $\SI{2}{mW}$ at $\SI{400}{nm}$. a) heatmap of the measured signal and b) heatmap of the simulated signal. c) transient signal at the position of the upper (blue) and lower (orange) polariton. The dots represent the measured signals, the solid lines the simulation.}
\label{fig:HeatmapsAndMaxima}
\vspace{2mm}
\end{figure}

In the following, we discuss the transient spectra of the same particles. The relative change in transmission $\Delta T/T_0$ after excitation with a $\SI{400}{nm}$ pump pulse at a pump power of $\SI{2}{mW}$ is presented in Figure\,\ref{fig:HeatmapsAndMaxima}a for various time delays. At the position of the static resonances, long lasting signal changes can be observed. The change at the transverse resonance can directly be attributed to the heating of the particles and effectively describes a broadening of the resonance. \cite{Stoll.2014} The two polariton resonances possess similar features with the same lifetime. On a first glance, two possible explanations seem appropriate. On the one hand, the excitation of polaritons or dark states has been suggested previously. \cite{Balci.2014, Peruffo.2022} On the other hand, the signal change directly translates to a widening of the peaks. Therefore, a higher damping of the resonances caused by particle heating has been proposed to be the source of the signal change also at the polariton resonances. \cite{Hao.2011, Simon.2016}

To calculate the influence of the transient heating, we inserted time dependent temperatures, obtained from the two-temperature model, into the simulation of the gold permittivity. This permittivity is then used in the Mie--Gans model. The initial conditions for solving eq \ref{eq:TwoTemperatureModel} are given by the electron and phonon temperatures at $t=0$. The lattice is initially at room temperature $T_\textrm{room}$ since the light is absorbed only by the electrons. Assuming that all absorbed light energy uniformly raises the temperature of the thermalized electron gas, \cite{Stoll.2014} the electron temperature at $t=0$ is \cite{delFatti.2006}
\begin{equation}
T_\textrm{e,0} = \sqrt{\frac{2 \Delta Q}{V \gamma}+T^2_\textrm{room}}
\label{eq:ElectronTempAfterExcitation}
\end{equation}
where $V$ represents the particle volume. The energy $\Delta Q$ deposited inside the particle is obtained from the laser fluence and the particle's absorption cross section. For an ensemble measurement in solution, determining $T_\textrm{e,0}$ is however not necessarily practicable due to variations in spot size and optical attenuation by the solution which lead to an uncertainty in $\Delta Q$. \cite{Park.2007} In a first simulation, we therefore leave the electron temperature after excitation variable. We will however later make use of eq \ref{eq:ElectronTempAfterExcitation} to compare the transient spectra of various pump fluences.

\begin{figure}[t]
    \centering
    \includegraphics{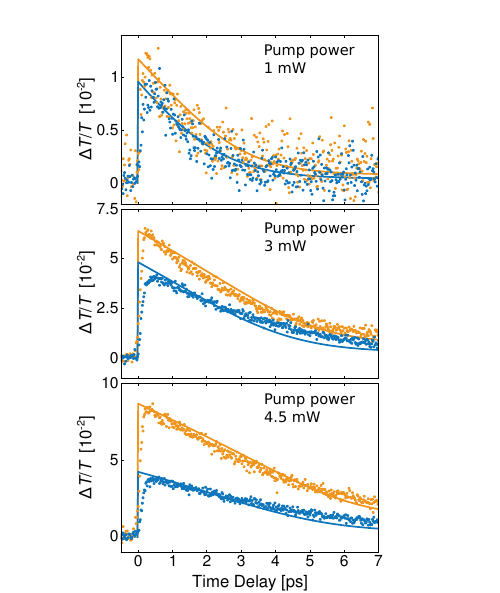}
    \caption{Measured and simulated transient signals at the polariton resonances of gold--TDBC core--shell nanorods pumped with $\SI{1}{mW}$, $\SI{3}{mW}$ and $\SI{4.5}{mW}$ (from top to bottom). Dots mark the experimental results while solid lines represent the simulations. The signal along the upper polariton indicated by blue marks, the signal along the lower polariton by orange ones.}
    \label{fig:TransientMaximaDifferentPumpPowers}
\end{figure}

Figure \ref{fig:HeatmapsAndMaxima} presents both measured and simulated data for a pump power of $\SI{2}{mW}$. For the simulation (Figure \ref{fig:HeatmapsAndMaxima}a), the only fitting parameter was $T_\textrm{e,0}$ (and a normalisation parameter). With $T_\textrm{e,0}=\SI{1000}{K}$, the model reproduces the measurement (Figure \ref{fig:HeatmapsAndMaxima}b) quite well. While the simulation closely resembles the measurements at the polariton resonances, it somewhat underestimates the magnitude at the transverse plasmon resonance, due to the underestimation of the transverse peaks as discussed earlier. The agreement can be verified by taking a closer look at the behaviour at the polariton wavelengths (Figure \ref{fig:HeatmapsAndMaxima}c). The simulation at these resonances agrees with the measurements reasonably well.

To further strengthen the hypothesis of a purely thermal origin of the transient signal as suspected in literature,\cite{Hao.2011, Simon.2016} we additionally investigated the temporal behaviour for pump powers of $\SI{1}{mW}$, $\SI{3}{mW}$ and $\SI{4.5}{mW}$. We can safely assume that the energy deposited inside a particle $\Delta Q$ is proportional to the pump power. Consequently, using $T_\textrm{e,0}=\SI{1000}{K}$ for $\SI{2}{mW}$ pump power, eq \ref{eq:ElectronTempAfterExcitation} predicts initial electron temperatures of $\SI{738}{K}$, $\SI{1206}{K}$ and $\SI{1462}{K}$ for the pump powers of $\SI{1}{mW}$, $\SI{3}{mW}$ and $\SI{4.5}{mW}$, respectively. 
Figure \ref{fig:TransientMaximaDifferentPumpPowers} presents the simulations using these initial temperatures in comparison to the measured data. The characteristic crossover from an exponential to an almost linear decay\cite{delFatti.2006} is nicely reproduced. 
Note that for the two higher pump powers, the ratio between the two signals (at upper and lower polariton position) was not ideal for the simulations. We therefore had to normalise the two signals separately to account for the higher measured change of the lower polariton. This discrepancy might be rooted in bleaching of the dye for higher pump powers which is not incorporated in our model.
Yet, this consistency of the model for different pump powers clearly supports the hypothesis of heat as the main source of transient signal also for strongly coupled nanoparticles. The high fluence transients can surely be fitted better if one allows for an adjustment of the various parameters entering the model. We think, however, that this consistent fit for  several fluences demonstrates that the gold heating is at the origin of the observed dynamics.

Although the signal change arises mainly at the polariton positions, it is unlikely that excited coupled states induce this change. On the one hand, the particles were pumped  with $\SI{400}{nm}$, a wavelength at which polaritons are not directly excited, on the other hand, the lifetime of such plasmon--exciton states is in the order of a few tens of femtoseconds, a timescale far below the lifetime of the signals in our measurements.


\section*{Conclusion} \vspace{-3.5mm}
In conclusion, we presented a method for the simulation of the full transient spectra of coupled plasmon-dye core-shell nanorods excited off the polariton resonance. To this end, we combined a well-established method for the transient description of bare gold nanoparticles with a method to simulate spectra of strongly coupled metal-dye particles that includes the dye saturation by the strong plasmon fields. The remarkable agreement between measured and simulated data in the present work confirms the suggestion of earlier studies that the ps-dynamics of these hybrid nanoparticles is dominated by transient heating effects.

\section*{Acknowledgements}\vspace{-3.5mm}
\noindent  We thank G\"unter Kewes for fruitful discussions. Additionally, F.S. acknowledges financial support by the DFG via the graduate school SALSA.

\bibliography{references.bib}

\begin{thebibliography}{37}%
\makeatletter
\providecommand \@ifxundefined [1]{%
 \@ifx{#1\undefined}
}%
\providecommand \@ifnum [1]{%
 \ifnum #1\expandafter \@firstoftwo
 \else \expandafter \@secondoftwo
 \fi
}%
\providecommand \@ifx [1]{%
 \ifx #1\expandafter \@firstoftwo
 \else \expandafter \@secondoftwo
 \fi
}%
\providecommand \natexlab [1]{#1}%
\providecommand \enquote  [1]{``#1''}%
\providecommand \bibnamefont  [1]{#1}%
\providecommand \bibfnamefont [1]{#1}%
\providecommand \citenamefont [1]{#1}%
\providecommand \href@noop [0]{\@secondoftwo}%
\providecommand \href [0]{\begingroup \@sanitize@url \@href}%
\providecommand \@href[1]{\@@startlink{#1}\@@href}%
\providecommand \@@href[1]{\endgroup#1\@@endlink}%
\providecommand \@sanitize@url [0]{\catcode `\\12\catcode `\$12\catcode
  `\&12\catcode `\#12\catcode `\^12\catcode `\_12\catcode `\%12\relax}%
\providecommand \@@startlink[1]{}%
\providecommand \@@endlink[0]{}%
\providecommand \url  [0]{\begingroup\@sanitize@url \@url }%
\providecommand \@url [1]{\endgroup\@href {#1}{\urlprefix }}%
\providecommand \urlprefix  [0]{URL }%
\providecommand \Eprint [0]{\href }%
\providecommand \doibase [0]{http://dx.doi.org/}%
\providecommand \selectlanguage [0]{\@gobble}%
\providecommand \bibinfo  [0]{\@secondoftwo}%
\providecommand \bibfield  [0]{\@secondoftwo}%
\providecommand \translation [1]{[#1]}%
\providecommand \BibitemOpen [0]{}%
\providecommand \bibitemStop [0]{}%
\providecommand \bibitemNoStop [0]{.\EOS\space}%
\providecommand \EOS [0]{\spacefactor3000\relax}%
\providecommand \BibitemShut  [1]{\csname bibitem#1\endcsname}%
\let\auto@bib@innerbib\@empty
\bibitem [{\citenamefont {Imamo\u{g}lu}\ \emph {et~al.}(1999)\citenamefont
  {Imamo\u{g}lu}, \citenamefont {Awschalom}, \citenamefont {Burkard},
  \citenamefont {DiVincenzo}, \citenamefont {Loss}, \citenamefont {Sherwin},\
  and\ \citenamefont {Small}}]{Imamoglu.1999}%
  \BibitemOpen
  \bibfield  {author} {\bibinfo {author} {\bibfnamefont {A.}~\bibnamefont
  {Imamo\u{g}lu}}, \bibinfo {author} {\bibfnamefont {D.~D.}\ \bibnamefont
  {Awschalom}}, \bibinfo {author} {\bibfnamefont {G.}~\bibnamefont {Burkard}},
  \bibinfo {author} {\bibfnamefont {D.~P.}\ \bibnamefont {DiVincenzo}},
  \bibinfo {author} {\bibfnamefont {D.}~\bibnamefont {Loss}}, \bibinfo {author}
  {\bibfnamefont {M.}~\bibnamefont {Sherwin}}, \ and\ \bibinfo {author}
  {\bibfnamefont {A.}~\bibnamefont {Small}},\ }\bibfield  {title} {\enquote
  {\bibinfo {title} {Quantum information processing using quantum dot spins and
  cavity qed},}\ }\href {\doibase 10.1103/PhysRevLett.83.4204} {\bibfield
  {journal} {\bibinfo  {journal} {Phys. Rev. Lett.}\ }\textbf {\bibinfo
  {volume} {83}},\ \bibinfo {pages} {4204--4207} (\bibinfo {year}
  {1999})}\BibitemShut {NoStop}%
\bibitem [{\citenamefont {Saba}\ \emph {et~al.}(2001)\citenamefont {Saba},
  \citenamefont {Ciuti}, \citenamefont {Bloch}, \citenamefont {Thierry-Mieg},
  \citenamefont {Andr{\'e}}, \citenamefont {Dang}, \citenamefont {Kundermann},
  \citenamefont {Mura}, \citenamefont {Bongiovanni}, \citenamefont {Staehli}
  \emph {et~al.}}]{Saba.2001}%
  \BibitemOpen
  \bibfield  {author} {\bibinfo {author} {\bibfnamefont {M.}~\bibnamefont
  {Saba}}, \bibinfo {author} {\bibfnamefont {C.}~\bibnamefont {Ciuti}},
  \bibinfo {author} {\bibfnamefont {J.}~\bibnamefont {Bloch}}, \bibinfo
  {author} {\bibfnamefont {V.}~\bibnamefont {Thierry-Mieg}}, \bibinfo {author}
  {\bibfnamefont {R.}~\bibnamefont {Andr{\'e}}}, \bibinfo {author}
  {\bibfnamefont {L.~S.}\ \bibnamefont {Dang}}, \bibinfo {author}
  {\bibfnamefont {S.}~\bibnamefont {Kundermann}}, \bibinfo {author}
  {\bibfnamefont {A.}~\bibnamefont {Mura}}, \bibinfo {author} {\bibfnamefont
  {G.}~\bibnamefont {Bongiovanni}}, \bibinfo {author} {\bibfnamefont
  {J.}~\bibnamefont {Staehli}},  \emph {et~al.},\ }\bibfield  {title} {\enquote
  {\bibinfo {title} {High-temperature ultrafast polariton parametric
  amplification in semiconductor microcavities},}\ }\href@noop {} {\bibfield
  {journal} {\bibinfo  {journal} {Nature}\ }\textbf {\bibinfo {volume} {414}},\
  \bibinfo {pages} {731} (\bibinfo {year} {2001})}\BibitemShut {NoStop}%
\bibitem [{\citenamefont {Hutchison}\ \emph {et~al.}(2012)\citenamefont
  {Hutchison}, \citenamefont {Schwartz}, \citenamefont {Genet}, \citenamefont
  {Devaux},\ and\ \citenamefont {Ebbesen}}]{Hutchison.2012}%
  \BibitemOpen
  \bibfield  {author} {\bibinfo {author} {\bibfnamefont {J.~A.}\ \bibnamefont
  {Hutchison}}, \bibinfo {author} {\bibfnamefont {T.}~\bibnamefont {Schwartz}},
  \bibinfo {author} {\bibfnamefont {C.}~\bibnamefont {Genet}}, \bibinfo
  {author} {\bibfnamefont {E.}~\bibnamefont {Devaux}}, \ and\ \bibinfo {author}
  {\bibfnamefont {T.~W.}\ \bibnamefont {Ebbesen}},\ }\bibfield  {title}
  {\enquote {\bibinfo {title} {Modifying chemical landscapes by coupling to
  vacuum fields},}\ }\href@noop {} {\bibfield  {journal} {\bibinfo  {journal}
  {Angew. Chem., Int. Ed.}\ }\textbf {\bibinfo {volume} {51}},\ \bibinfo
  {pages} {1592--1596} (\bibinfo {year} {2012})}\BibitemShut {NoStop}%
\bibitem [{\citenamefont {T{\"o}rm{\"a}}\ and\ \citenamefont
  {Barnes}(2014)}]{Torma.2014}%
  \BibitemOpen
  \bibfield  {author} {\bibinfo {author} {\bibfnamefont {P.}~\bibnamefont
  {T{\"o}rm{\"a}}}\ and\ \bibinfo {author} {\bibfnamefont {W.~L.}\ \bibnamefont
  {Barnes}},\ }\bibfield  {title} {\enquote {\bibinfo {title} {Strong coupling
  between surface plasmon polaritons and emitters: A review},}\ }\href
  {\doibase 10.1088/0034-4885/78/1/013901} {\bibfield  {journal} {\bibinfo
  {journal} {Rep. Prog. Phys.}\ }\textbf {\bibinfo {volume} {78}},\ \bibinfo
  {pages} {013901} (\bibinfo {year} {2014})}\BibitemShut {NoStop}%
\bibitem [{\citenamefont {Zheng}\ \emph {et~al.}(2017)\citenamefont {Zheng},
  \citenamefont {Zhang}, \citenamefont {Deng}, \citenamefont {Kang},
  \citenamefont {Nordlander},\ and\ \citenamefont {Xu}}]{Zheng.2017}%
  \BibitemOpen
  \bibfield  {author} {\bibinfo {author} {\bibfnamefont {D.}~\bibnamefont
  {Zheng}}, \bibinfo {author} {\bibfnamefont {S.}~\bibnamefont {Zhang}},
  \bibinfo {author} {\bibfnamefont {Q.}~\bibnamefont {Deng}}, \bibinfo {author}
  {\bibfnamefont {M.}~\bibnamefont {Kang}}, \bibinfo {author} {\bibfnamefont
  {P.}~\bibnamefont {Nordlander}}, \ and\ \bibinfo {author} {\bibfnamefont
  {H.}~\bibnamefont {Xu}},\ }\bibfield  {title} {\enquote {\bibinfo {title}
  {Manipulating coherent plasmon–exciton interaction in a single silver
  nanorod on monolayer wse2},}\ }\href {\doibase 10.1021/acs.nanolett.7b01176}
  {\bibfield  {journal} {\bibinfo  {journal} {Nano Lett.}\ }\textbf {\bibinfo
  {volume} {17}},\ \bibinfo {pages} {3809--3814} (\bibinfo {year}
  {2017})}\BibitemShut {NoStop}%
\bibitem [{\citenamefont {Zengin}\ \emph {et~al.}(2015)\citenamefont {Zengin},
  \citenamefont {Wers{\"a}ll}, \citenamefont {Nilsson}, \citenamefont
  {Antosiewicz}, \citenamefont {K{\"a}ll},\ and\ \citenamefont
  {Shegai}}]{Zengin.2015}%
  \BibitemOpen
  \bibfield  {author} {\bibinfo {author} {\bibfnamefont {G.}~\bibnamefont
  {Zengin}}, \bibinfo {author} {\bibfnamefont {M.}~\bibnamefont {Wers{\"a}ll}},
  \bibinfo {author} {\bibfnamefont {S.}~\bibnamefont {Nilsson}}, \bibinfo
  {author} {\bibfnamefont {T.~J.}\ \bibnamefont {Antosiewicz}}, \bibinfo
  {author} {\bibfnamefont {M.}~\bibnamefont {K{\"a}ll}}, \ and\ \bibinfo
  {author} {\bibfnamefont {T.}~\bibnamefont {Shegai}},\ }\bibfield  {title}
  {\enquote {\bibinfo {title} {{Realizing Strong Light--Matter Interactions
  between Single-Nanoparticle Plasmons and Molecular Excitons at Ambient
  Conditions}},}\ }\href {\doibase 10.1103/PhysRevLett.114.157401} {\bibfield
  {journal} {\bibinfo  {journal} {{Phys. Rev. Lett.}}\ }\textbf {\bibinfo
  {volume} {114}},\ \bibinfo {pages} {157401} (\bibinfo {year}
  {2015})}\BibitemShut {NoStop}%
\bibitem [{\citenamefont {Zhou}\ \emph {et~al.}(2016)\citenamefont {Zhou},
  \citenamefont {Yuan}, \citenamefont {Gao}, \citenamefont {Li},\ and\
  \citenamefont {Yang}}]{Zhou.2016}%
  \BibitemOpen
  \bibfield  {author} {\bibinfo {author} {\bibfnamefont {N.}~\bibnamefont
  {Zhou}}, \bibinfo {author} {\bibfnamefont {M.}~\bibnamefont {Yuan}}, \bibinfo
  {author} {\bibfnamefont {Y.}~\bibnamefont {Gao}}, \bibinfo {author}
  {\bibfnamefont {D.}~\bibnamefont {Li}}, \ and\ \bibinfo {author}
  {\bibfnamefont {D.}~\bibnamefont {Yang}},\ }\bibfield  {title} {\enquote
  {\bibinfo {title} {Silver nanoshell plasmonically controlled emission of
  semiconductor quantum dots in the strong coupling regime},}\ }\href {\doibase
  10.1021/acsnano.5b07400} {\bibfield  {journal} {\bibinfo  {journal} {ACS
  Nano}\ }\textbf {\bibinfo {volume} {10}},\ \bibinfo {pages} {4154--4163}
  (\bibinfo {year} {2016})}\BibitemShut {NoStop}%
\bibitem [{\citenamefont {Melnikau}\ \emph {et~al.}(2016)\citenamefont
  {Melnikau}, \citenamefont {Esteban}, \citenamefont {Savateeva}, \citenamefont
  {S\'anchez-Iglesias}, \citenamefont {Grzelczak}, \citenamefont {Schmidt},
  \citenamefont {Liz-Marz\'an}, \citenamefont {Aizpurua},\ and\ \citenamefont
  {Rakovich}}]{Melnikau.2016}%
  \BibitemOpen
  \bibfield  {author} {\bibinfo {author} {\bibfnamefont {D.}~\bibnamefont
  {Melnikau}}, \bibinfo {author} {\bibfnamefont {R.}~\bibnamefont {Esteban}},
  \bibinfo {author} {\bibfnamefont {D.}~\bibnamefont {Savateeva}}, \bibinfo
  {author} {\bibfnamefont {A.}~\bibnamefont {S\'anchez-Iglesias}}, \bibinfo
  {author} {\bibfnamefont {M.}~\bibnamefont {Grzelczak}}, \bibinfo {author}
  {\bibfnamefont {M.~K.}\ \bibnamefont {Schmidt}}, \bibinfo {author}
  {\bibfnamefont {L.~M.}\ \bibnamefont {Liz-Marz\'an}}, \bibinfo {author}
  {\bibfnamefont {J.}~\bibnamefont {Aizpurua}}, \ and\ \bibinfo {author}
  {\bibfnamefont {Y.~P.}\ \bibnamefont {Rakovich}},\ }\bibfield  {title}
  {\enquote {\bibinfo {title} {Rabi splitting in photoluminescence spectra of
  hybrid systems of gold nanorods and j-aggregates},}\ }\href {\doibase
  10.1021/acs.jpclett.5b02512} {\bibfield  {journal} {\bibinfo  {journal} {J.
  Phys. Chem. Lett.}\ }\textbf {\bibinfo {volume} {7}},\ \bibinfo {pages}
  {354--362} (\bibinfo {year} {2016})}\BibitemShut {NoStop}%
\bibitem [{\citenamefont {Eizner}\ \emph {et~al.}(2015)\citenamefont {Eizner},
  \citenamefont {Avayu}, \citenamefont {Ditcovski},\ and\ \citenamefont
  {Ellenbogen}}]{Eizner.2015}%
  \BibitemOpen
  \bibfield  {author} {\bibinfo {author} {\bibfnamefont {E.}~\bibnamefont
  {Eizner}}, \bibinfo {author} {\bibfnamefont {O.}~\bibnamefont {Avayu}},
  \bibinfo {author} {\bibfnamefont {R.}~\bibnamefont {Ditcovski}}, \ and\
  \bibinfo {author} {\bibfnamefont {T.}~\bibnamefont {Ellenbogen}},\ }\bibfield
   {title} {\enquote {\bibinfo {title} {Aluminum nanoantenna complexes for
  strong coupling between excitons and localized surface plasmons},}\ }\href
  {\doibase 10.1021/acs.nanolett.5b02584} {\bibfield  {journal} {\bibinfo
  {journal} {Nano Lett.}\ }\textbf {\bibinfo {volume} {15}},\ \bibinfo {pages}
  {6215--6221} (\bibinfo {year} {2015})}\BibitemShut {NoStop}%
\bibitem [{\citenamefont {Vasa}\ \emph {et~al.}(2013)\citenamefont {Vasa},
  \citenamefont {Wang}, \citenamefont {Pomraenke}, \citenamefont {Lammers},
  \citenamefont {Maiuri}, \citenamefont {Manzoni}, \citenamefont {Cerullo},\
  and\ \citenamefont {Lienau}}]{Vasa.2013}%
  \BibitemOpen
  \bibfield  {author} {\bibinfo {author} {\bibfnamefont {P.}~\bibnamefont
  {Vasa}}, \bibinfo {author} {\bibfnamefont {W.}~\bibnamefont {Wang}}, \bibinfo
  {author} {\bibfnamefont {R.}~\bibnamefont {Pomraenke}}, \bibinfo {author}
  {\bibfnamefont {M.}~\bibnamefont {Lammers}}, \bibinfo {author} {\bibfnamefont
  {M.}~\bibnamefont {Maiuri}}, \bibinfo {author} {\bibfnamefont
  {C.}~\bibnamefont {Manzoni}}, \bibinfo {author} {\bibfnamefont
  {G.}~\bibnamefont {Cerullo}}, \ and\ \bibinfo {author} {\bibfnamefont
  {C.}~\bibnamefont {Lienau}},\ }\bibfield  {title} {\enquote {\bibinfo {title}
  {{Real-Time Observation of Ultrafast Rabi Oscillations between Excitons and
  Plasmons in Metal Nanostructures with J-Aggregates}},}\ }\href {\doibase
  10.1038/nphoton.2012.340} {\bibfield  {journal} {\bibinfo  {journal} {{Nat.
  Photonics}}\ }\textbf {\bibinfo {volume} {7}},\ \bibinfo {pages} {128--132}
  (\bibinfo {year} {2013})}\BibitemShut {NoStop}%
\bibitem [{\citenamefont {Peruffo}, \citenamefont {Mancin},\ and\ \citenamefont
  {Collini}(2023)}]{Peruffo.2023}%
  \BibitemOpen
  \bibfield  {author} {\bibinfo {author} {\bibfnamefont {N.}~\bibnamefont
  {Peruffo}}, \bibinfo {author} {\bibfnamefont {F.}~\bibnamefont {Mancin}}, \
  and\ \bibinfo {author} {\bibfnamefont {E.}~\bibnamefont {Collini}},\
  }\bibfield  {title} {\enquote {\bibinfo {title} {{Coherent Dynamics in
  Solutions of Colloidal Plexcitonic Nanohybrids at Room Temperature}},}\
  }\href@noop {} {\bibfield  {journal} {\bibinfo  {journal} {Adv. Opt. Mater.}\
  ,\ \bibinfo {pages} {2203010}} (\bibinfo {year} {2023})}\BibitemShut
  {NoStop}%
\bibitem [{\citenamefont {Hao}\ \emph {et~al.}(2011)\citenamefont {Hao},
  \citenamefont {Wang}, \citenamefont {Jiang}, \citenamefont {Chen},
  \citenamefont {Ueno}, \citenamefont {Wang}, \citenamefont {Misawa},\ and\
  \citenamefont {Sun}}]{Hao.2011}%
  \BibitemOpen
  \bibfield  {author} {\bibinfo {author} {\bibfnamefont {Y.-W.}\ \bibnamefont
  {Hao}}, \bibinfo {author} {\bibfnamefont {H.-Y.}\ \bibnamefont {Wang}},
  \bibinfo {author} {\bibfnamefont {Y.}~\bibnamefont {Jiang}}, \bibinfo
  {author} {\bibfnamefont {Q.-D.}\ \bibnamefont {Chen}}, \bibinfo {author}
  {\bibfnamefont {K.}~\bibnamefont {Ueno}}, \bibinfo {author} {\bibfnamefont
  {W.-Q.}\ \bibnamefont {Wang}}, \bibinfo {author} {\bibfnamefont
  {H.}~\bibnamefont {Misawa}}, \ and\ \bibinfo {author} {\bibfnamefont {H.-B.}\
  \bibnamefont {Sun}},\ }\bibfield  {title} {\enquote {\bibinfo {title}
  {Hybrid-state dynamics of gold nanorods/dye j-aggregates under strong
  coupling},}\ }\href {\doibase 10.1002/ange.201101699} {\bibfield  {journal}
  {\bibinfo  {journal} {Angew. Chem. Int. Ed.}\ }\textbf {\bibinfo {volume}
  {123}},\ \bibinfo {pages} {7970--7974} (\bibinfo {year} {2011})}\BibitemShut
  {NoStop}%
\bibitem [{\citenamefont {Simon}\ \emph {et~al.}(2016)\citenamefont {Simon},
  \citenamefont {Melnikau}, \citenamefont {S\'anchez-Iglesias}, \citenamefont
  {Grzelczak}, \citenamefont {Liz-Marz\'an}, \citenamefont {Rakovich},
  \citenamefont {Feldmann},\ and\ \citenamefont {Urban}}]{Simon.2016}%
  \BibitemOpen
  \bibfield  {author} {\bibinfo {author} {\bibfnamefont {T.}~\bibnamefont
  {Simon}}, \bibinfo {author} {\bibfnamefont {D.}~\bibnamefont {Melnikau}},
  \bibinfo {author} {\bibfnamefont {A.}~\bibnamefont {S\'anchez-Iglesias}},
  \bibinfo {author} {\bibfnamefont {M.}~\bibnamefont {Grzelczak}}, \bibinfo
  {author} {\bibfnamefont {L.~M.}\ \bibnamefont {Liz-Marz\'an}}, \bibinfo
  {author} {\bibfnamefont {Y.}~\bibnamefont {Rakovich}}, \bibinfo {author}
  {\bibfnamefont {J.}~\bibnamefont {Feldmann}}, \ and\ \bibinfo {author}
  {\bibfnamefont {A.~S.}\ \bibnamefont {Urban}},\ }\bibfield  {title} {\enquote
  {\bibinfo {title} {Exploring the optical nonlinearities of plasmon-exciton
  hybrid resonances in coupled colloidal nanostructures},}\ }\href {\doibase
  10.1021/acs.jpcc.6b04658} {\bibfield  {journal} {\bibinfo  {journal} {J.
  Phys. Chem. C}\ }\textbf {\bibinfo {volume} {120}},\ \bibinfo {pages}
  {12226--12233} (\bibinfo {year} {2016})}\BibitemShut {NoStop}%
\bibitem [{\citenamefont {Eizner}\ \emph {et~al.}(2017)\citenamefont {Eizner},
  \citenamefont {Akulov}, \citenamefont {Schwartz},\ and\ \citenamefont
  {Ellenbogen}}]{Eizner.2017}%
  \BibitemOpen
  \bibfield  {author} {\bibinfo {author} {\bibfnamefont {E.}~\bibnamefont
  {Eizner}}, \bibinfo {author} {\bibfnamefont {K.}~\bibnamefont {Akulov}},
  \bibinfo {author} {\bibfnamefont {T.}~\bibnamefont {Schwartz}}, \ and\
  \bibinfo {author} {\bibfnamefont {T.}~\bibnamefont {Ellenbogen}},\ }\bibfield
   {title} {\enquote {\bibinfo {title} {{Temporal Dynamics of Localized
  Exciton--Polaritons in Composite Organic--Plasmonic Metasurfaces}},}\ }\href
  {\doibase 10.1021/acs.nanolett.7b03751} {\bibfield  {journal} {\bibinfo
  {journal} {Nano Lett.}\ }\textbf {\bibinfo {volume} {17}},\ \bibinfo {pages}
  {7675--7683} (\bibinfo {year} {2017})}\BibitemShut {NoStop}%
\bibitem [{\citenamefont {Finkelstein-Shapiro}\ \emph
  {et~al.}(2021)\citenamefont {Finkelstein-Shapiro}, \citenamefont {Mante},
  \citenamefont {Sarisozen}, \citenamefont {Wittenbecher}, \citenamefont
  {Minda}, \citenamefont {Balci}, \citenamefont {Pullerits},\ and\
  \citenamefont {Zigmantas}}]{Finkelstein-Shapiro.2021}%
  \BibitemOpen
  \bibfield  {author} {\bibinfo {author} {\bibfnamefont {D.}~\bibnamefont
  {Finkelstein-Shapiro}}, \bibinfo {author} {\bibfnamefont {P.-A.}\
  \bibnamefont {Mante}}, \bibinfo {author} {\bibfnamefont {S.}~\bibnamefont
  {Sarisozen}}, \bibinfo {author} {\bibfnamefont {L.}~\bibnamefont
  {Wittenbecher}}, \bibinfo {author} {\bibfnamefont {I.}~\bibnamefont {Minda}},
  \bibinfo {author} {\bibfnamefont {S.}~\bibnamefont {Balci}}, \bibinfo
  {author} {\bibfnamefont {T.}~\bibnamefont {Pullerits}}, \ and\ \bibinfo
  {author} {\bibfnamefont {D.}~\bibnamefont {Zigmantas}},\ }\bibfield  {title}
  {\enquote {\bibinfo {title} {Understanding radiative transitions and
  relaxation pathways in plexcitons},}\ }\href@noop {} {\bibfield  {journal}
  {\bibinfo  {journal} {Chem}\ }\textbf {\bibinfo {volume} {7}},\ \bibinfo
  {pages} {1092--1107} (\bibinfo {year} {2021})}\BibitemShut {NoStop}%
\bibitem [{\citenamefont {Voisin}\ \emph {et~al.}(2001)\citenamefont {Voisin},
  \citenamefont {Del~Fatti}, \citenamefont {Christofilos},\ and\ \citenamefont
  {Vall{\'e}e}}]{Voisin.2001}%
  \BibitemOpen
  \bibfield  {author} {\bibinfo {author} {\bibfnamefont {C.}~\bibnamefont
  {Voisin}}, \bibinfo {author} {\bibfnamefont {N.}~\bibnamefont {Del~Fatti}},
  \bibinfo {author} {\bibfnamefont {D.}~\bibnamefont {Christofilos}}, \ and\
  \bibinfo {author} {\bibfnamefont {F.}~\bibnamefont {Vall{\'e}e}},\ }\bibfield
   {title} {\enquote {\bibinfo {title} {Ultrafast electron dynamics and optical
  nonlinearities in metal nanoparticles},}\ }\href {\doibase 10.1021/jp0038153}
  {\bibfield  {journal} {\bibinfo  {journal} {J. Phys. Chem. B}\ }\textbf
  {\bibinfo {volume} {105}},\ \bibinfo {pages} {2264--2280} (\bibinfo {year}
  {2001})}\BibitemShut {NoStop}%
\bibitem [{\citenamefont {Hodak}, \citenamefont {Martini},\ and\ \citenamefont
  {Hartland}(1998)}]{Hodak.1998}%
  \BibitemOpen
  \bibfield  {author} {\bibinfo {author} {\bibfnamefont {J.~H.}\ \bibnamefont
  {Hodak}}, \bibinfo {author} {\bibfnamefont {I.}~\bibnamefont {Martini}}, \
  and\ \bibinfo {author} {\bibfnamefont {G.~V.}\ \bibnamefont {Hartland}},\
  }\bibfield  {title} {\enquote {\bibinfo {title} {{Spectroscopy and Dynamics
  of Nanometer-Sized Noble Metal Particles}},}\ }\href@noop {} {\bibfield
  {journal} {\bibinfo  {journal} {J. Phys. Chem. B}\ }\textbf {\bibinfo
  {volume} {102}},\ \bibinfo {pages} {6958--6967} (\bibinfo {year}
  {1998})}\BibitemShut {NoStop}%
\bibitem [{\citenamefont {Park}\ \emph {et~al.}(2007)\citenamefont {Park},
  \citenamefont {Pelton}, \citenamefont {Liu}, \citenamefont {Guyot-Sionnest},\
  and\ \citenamefont {Scherer}}]{Park.2007}%
  \BibitemOpen
  \bibfield  {author} {\bibinfo {author} {\bibfnamefont {S.}~\bibnamefont
  {Park}}, \bibinfo {author} {\bibfnamefont {M.}~\bibnamefont {Pelton}},
  \bibinfo {author} {\bibfnamefont {M.}~\bibnamefont {Liu}}, \bibinfo {author}
  {\bibfnamefont {P.}~\bibnamefont {Guyot-Sionnest}}, \ and\ \bibinfo {author}
  {\bibfnamefont {N.~F.}\ \bibnamefont {Scherer}},\ }\bibfield  {title}
  {\enquote {\bibinfo {title} {Ultrafast resonant dynamics of surface plasmons
  in gold nanorods},}\ }\href@noop {} {\bibfield  {journal} {\bibinfo
  {journal} {J. Phys. Chem. C}\ }\textbf {\bibinfo {volume} {111}},\ \bibinfo
  {pages} {116--123} (\bibinfo {year} {2007})}\BibitemShut {NoStop}%
\bibitem [{\citenamefont {Stoll}\ \emph {et~al.}(2014)\citenamefont {Stoll},
  \citenamefont {Maioli}, \citenamefont {Crut}, \citenamefont {Del~Fatti},\
  and\ \citenamefont {Vall{\'e}e}}]{Stoll.2014}%
  \BibitemOpen
  \bibfield  {author} {\bibinfo {author} {\bibfnamefont {T.}~\bibnamefont
  {Stoll}}, \bibinfo {author} {\bibfnamefont {P.}~\bibnamefont {Maioli}},
  \bibinfo {author} {\bibfnamefont {A.}~\bibnamefont {Crut}}, \bibinfo {author}
  {\bibfnamefont {N.}~\bibnamefont {Del~Fatti}}, \ and\ \bibinfo {author}
  {\bibfnamefont {F.}~\bibnamefont {Vall{\'e}e}},\ }\bibfield  {title}
  {\enquote {\bibinfo {title} {{Advances in femto-nano-optics: ultrafast
  nonlinearity of metal nanoparticles}},}\ }\href {\doibase
  10.1140/epjb/e2014-50515-4} {\bibfield  {journal} {\bibinfo  {journal} {Eur.
  Phys. J. B}\ }\textbf {\bibinfo {volume} {87}},\ \bibinfo {pages} {260}
  (\bibinfo {year} {2014})}\BibitemShut {NoStop}%
\bibitem [{\citenamefont {Stete}\ \emph {et~al.}(2022)\citenamefont {Stete},
  \citenamefont {Koopman}, \citenamefont {Henkel}, \citenamefont {Benson},
  \citenamefont {Kewes},\ and\ \citenamefont {Bargheer}}]{Stete.2020}%
  \BibitemOpen
  \bibfield  {author} {\bibinfo {author} {\bibfnamefont {F.}~\bibnamefont
  {Stete}}, \bibinfo {author} {\bibfnamefont {W.}~\bibnamefont {Koopman}},
  \bibinfo {author} {\bibfnamefont {C.}~\bibnamefont {Henkel}}, \bibinfo
  {author} {\bibfnamefont {O.}~\bibnamefont {Benson}}, \bibinfo {author}
  {\bibfnamefont {G.}~\bibnamefont {Kewes}}, \ and\ \bibinfo {author}
  {\bibfnamefont {M.}~\bibnamefont {Bargheer}},\ }\bibfield  {title} {\enquote
  {\bibinfo {title} {Vacuum-induced saturation in plasmonic nanoparticles},}\
  }\href@noop {} {\bibfield  {journal} {\bibinfo  {journal} {chemrxiv preprint
  10.26434/chemrxiv-2022-r6qc2}\ ,\ \bibinfo {pages} {in press at ACS
  Photonics}} (\bibinfo {year} {2022})}\BibitemShut {NoStop}%
\bibitem [{\citenamefont {Stete}, \citenamefont {Koopman},\ and\ \citenamefont
  {Bargheer}(2017)}]{Stete.2017}%
  \BibitemOpen
  \bibfield  {author} {\bibinfo {author} {\bibfnamefont {F.}~\bibnamefont
  {Stete}}, \bibinfo {author} {\bibfnamefont {W.}~\bibnamefont {Koopman}}, \
  and\ \bibinfo {author} {\bibfnamefont {M.}~\bibnamefont {Bargheer}},\
  }\bibfield  {title} {\enquote {\bibinfo {title} {Signatures of strong
  coupling on nanoparticles: Revealing absorption anticrossing by tuning the
  dielectric environment},}\ }\href {\doibase 10.1021/acsphotonics.7b00113}
  {\bibfield  {journal} {\bibinfo  {journal} {ACS Photonics}\ }\textbf
  {\bibinfo {volume} {4}},\ \bibinfo {pages} {1669--1676} (\bibinfo {year}
  {2017})}\BibitemShut {NoStop}%
\bibitem [{\citenamefont {Stete}\ \emph {et~al.}(2018)\citenamefont {Stete},
  \citenamefont {Scho{\ss}au}, \citenamefont {Bargheer},\ and\ \citenamefont
  {Koopman}}]{Stete.2018}%
  \BibitemOpen
  \bibfield  {author} {\bibinfo {author} {\bibfnamefont {F.}~\bibnamefont
  {Stete}}, \bibinfo {author} {\bibfnamefont {P.}~\bibnamefont {Scho{\ss}au}},
  \bibinfo {author} {\bibfnamefont {M.}~\bibnamefont {Bargheer}}, \ and\
  \bibinfo {author} {\bibfnamefont {W.}~\bibnamefont {Koopman}},\ }\bibfield
  {title} {\enquote {\bibinfo {title} {Size-dependent coupling of hybrid
  core--shell nanorods: Toward single-emitter strong-coupling},}\ }\href
  {\doibase 10.1021/acs.jpcc.8b04204} {\bibfield  {journal} {\bibinfo
  {journal} {J. Phys. Chem. C}\ }\textbf {\bibinfo {volume} {122}},\ \bibinfo
  {pages} {17976--17982} (\bibinfo {year} {2018})}\BibitemShut {NoStop}%
\bibitem [{\citenamefont {Rosei}(1974)}]{Rosei.1974}%
  \BibitemOpen
  \bibfield  {author} {\bibinfo {author} {\bibfnamefont {R.}~\bibnamefont
  {Rosei}},\ }\bibfield  {title} {\enquote {\bibinfo {title} {Temperature
  modulation of the optical transitions involving the fermi surface in ag:
  Theory},}\ }\href {\doibase 10.1103/PhysRevB.10.474} {\bibfield  {journal}
  {\bibinfo  {journal} {Phys. Rev. B}\ }\textbf {\bibinfo {volume} {10}},\
  \bibinfo {pages} {474} (\bibinfo {year} {1974})}\BibitemShut {NoStop}%
\bibitem [{\citenamefont {Rosei}, \citenamefont {Antonangeli},\ and\
  \citenamefont {Grassano}(1973)}]{Rosei.1973}%
  \BibitemOpen
  \bibfield  {author} {\bibinfo {author} {\bibfnamefont {R.}~\bibnamefont
  {Rosei}}, \bibinfo {author} {\bibfnamefont {F.}~\bibnamefont {Antonangeli}},
  \ and\ \bibinfo {author} {\bibfnamefont {U.}~\bibnamefont {Grassano}},\
  }\bibfield  {title} {\enquote {\bibinfo {title} {D bands position and width
  in gold from very low temperature thermomodulation measurements},}\ }\href
  {\doibase 10.1016/0039-6028(73)90359-2} {\bibfield  {journal} {\bibinfo
  {journal} {Surf. Sci.}\ }\textbf {\bibinfo {volume} {37}},\ \bibinfo {pages}
  {689--699} (\bibinfo {year} {1973})}\BibitemShut {NoStop}%
\bibitem [{\citenamefont {Guerrisi}, \citenamefont {Rosei},\ and\ \citenamefont
  {Winsemius}(1975)}]{Guerrisi.1975}%
  \BibitemOpen
  \bibfield  {author} {\bibinfo {author} {\bibfnamefont {M.}~\bibnamefont
  {Guerrisi}}, \bibinfo {author} {\bibfnamefont {R.}~\bibnamefont {Rosei}}, \
  and\ \bibinfo {author} {\bibfnamefont {P.}~\bibnamefont {Winsemius}},\
  }\bibfield  {title} {\enquote {\bibinfo {title} {Splitting of the interband
  absorption edge in au},}\ }\href {\doibase 10.1103/PhysRevB.12.557}
  {\bibfield  {journal} {\bibinfo  {journal} {Phys. Rev. B}\ }\textbf {\bibinfo
  {volume} {12}},\ \bibinfo {pages} {557--563} (\bibinfo {year}
  {1975})}\BibitemShut {NoStop}%
\bibitem [{\citenamefont {Winsemius}, \citenamefont {Guerrisi},\ and\
  \citenamefont {Rosei}(1975)}]{Winsemius.1975}%
  \BibitemOpen
  \bibfield  {author} {\bibinfo {author} {\bibfnamefont {P.}~\bibnamefont
  {Winsemius}}, \bibinfo {author} {\bibfnamefont {M.}~\bibnamefont {Guerrisi}},
  \ and\ \bibinfo {author} {\bibfnamefont {R.}~\bibnamefont {Rosei}},\
  }\bibfield  {title} {\enquote {\bibinfo {title} {Splitting of the interband
  absorption edge in au: Temperature dependence},}\ }\href {\doibase
  10.1103/PhysRevB.12.4570} {\bibfield  {journal} {\bibinfo  {journal} {Phys.
  Rev. B}\ }\textbf {\bibinfo {volume} {12}},\ \bibinfo {pages} {4570--4572}
  (\bibinfo {year} {1975})}\BibitemShut {NoStop}%
\bibitem [{\citenamefont {Bohren}\ and\ \citenamefont
  {Huffman}(2008)}]{Bohren.2008}%
  \BibitemOpen
  \bibfield  {author} {\bibinfo {author} {\bibfnamefont {C.~F.}\ \bibnamefont
  {Bohren}}\ and\ \bibinfo {author} {\bibfnamefont {D.~R.}\ \bibnamefont
  {Huffman}},\ }\href@noop {} {\emph {\bibinfo {title} {Absorption and
  Scattering of Light by Small Particles}}}\ (\bibinfo  {publisher} {John Wiley
  \& Sons},\ \bibinfo {year} {2008})\BibitemShut {NoStop}%
\bibitem [{\citenamefont {Zengin}\ \emph {et~al.}(2013)\citenamefont {Zengin},
  \citenamefont {Johansson}, \citenamefont {Johansson}, \citenamefont
  {Antosiewicz}, \citenamefont {K{\"a}ll},\ and\ \citenamefont
  {Shegai}}]{Zengin.2013}%
  \BibitemOpen
  \bibfield  {author} {\bibinfo {author} {\bibfnamefont {G.}~\bibnamefont
  {Zengin}}, \bibinfo {author} {\bibfnamefont {G.}~\bibnamefont {Johansson}},
  \bibinfo {author} {\bibfnamefont {P.}~\bibnamefont {Johansson}}, \bibinfo
  {author} {\bibfnamefont {T.~J.}\ \bibnamefont {Antosiewicz}}, \bibinfo
  {author} {\bibfnamefont {M.}~\bibnamefont {K{\"a}ll}}, \ and\ \bibinfo
  {author} {\bibfnamefont {T.}~\bibnamefont {Shegai}},\ }\bibfield  {title}
  {\enquote {\bibinfo {title} {Approaching the strong coupling limit in single
  plasmonic nanorods interacting with j-aggregates},}\ }\href@noop {}
  {\bibfield  {journal} {\bibinfo  {journal} {Sci. Rep.}\ }\textbf {\bibinfo
  {volume} {3}},\ \bibinfo {pages} {3074} (\bibinfo {year} {2013})}\BibitemShut
  {NoStop}%
\bibitem [{\citenamefont {Kelly}\ \emph {et~al.}(2003)\citenamefont {Kelly},
  \citenamefont {Coronado}, \citenamefont {Zhao},\ and\ \citenamefont
  {Schatz}}]{Kelly.2003}%
  \BibitemOpen
  \bibfield  {author} {\bibinfo {author} {\bibfnamefont {K.~L.}\ \bibnamefont
  {Kelly}}, \bibinfo {author} {\bibfnamefont {E.}~\bibnamefont {Coronado}},
  \bibinfo {author} {\bibfnamefont {L.~L.}\ \bibnamefont {Zhao}}, \ and\
  \bibinfo {author} {\bibfnamefont {G.~C.}\ \bibnamefont {Schatz}},\ }\bibfield
   {title} {\enquote {\bibinfo {title} {The optical properties of metal
  nanoparticles: The influence of size, shape, and dielectric environment},}\
  }\href {\doibase 10.1021/jp026731y} {\bibfield  {journal} {\bibinfo
  {journal} {J. Phys. Chem. B}\ }\textbf {\bibinfo {volume} {107}},\ \bibinfo
  {pages} {668--677} (\bibinfo {year} {2003})}\BibitemShut {NoStop}%
\bibitem [{\citenamefont {Olmon}\ \emph {et~al.}(2012)\citenamefont {Olmon},
  \citenamefont {Slovick}, \citenamefont {Johnson}, \citenamefont {Shelton},
  \citenamefont {Oh}, \citenamefont {Boreman},\ and\ \citenamefont
  {Raschke}}]{Olmon.2012}%
  \BibitemOpen
  \bibfield  {author} {\bibinfo {author} {\bibfnamefont {R.~L.}\ \bibnamefont
  {Olmon}}, \bibinfo {author} {\bibfnamefont {B.}~\bibnamefont {Slovick}},
  \bibinfo {author} {\bibfnamefont {T.~W.}\ \bibnamefont {Johnson}}, \bibinfo
  {author} {\bibfnamefont {D.}~\bibnamefont {Shelton}}, \bibinfo {author}
  {\bibfnamefont {S.-H.}\ \bibnamefont {Oh}}, \bibinfo {author} {\bibfnamefont
  {G.~D.}\ \bibnamefont {Boreman}}, \ and\ \bibinfo {author} {\bibfnamefont
  {M.~B.}\ \bibnamefont {Raschke}},\ }\bibfield  {title} {\enquote {\bibinfo
  {title} {{Optical Dielectric Function of Gold}},}\ }\href {\doibase
  10.1103/PhysRevB.86.235147} {\bibfield  {journal} {\bibinfo  {journal} {Phys.
  Rev. B}\ }\textbf {\bibinfo {volume} {86}},\ \bibinfo {pages} {235147}
  (\bibinfo {year} {2012})}\BibitemShut {NoStop}%
\bibitem [{\citenamefont {Rethfeld}\ \emph {et~al.}(2017)\citenamefont
  {Rethfeld}, \citenamefont {Ivanov}, \citenamefont {Garcia},\ and\
  \citenamefont {Anisimov}}]{Rethfeld.2017}%
  \BibitemOpen
  \bibfield  {author} {\bibinfo {author} {\bibfnamefont {B.}~\bibnamefont
  {Rethfeld}}, \bibinfo {author} {\bibfnamefont {D.~S.}\ \bibnamefont
  {Ivanov}}, \bibinfo {author} {\bibfnamefont {M.~E.}\ \bibnamefont {Garcia}},
  \ and\ \bibinfo {author} {\bibfnamefont {S.~I.}\ \bibnamefont {Anisimov}},\
  }\bibfield  {title} {\enquote {\bibinfo {title} {Modelling ultrafast laser
  ablation},}\ }\href {\doibase 10.1088/1361-6463/50/19/193001} {\bibfield
  {journal} {\bibinfo  {journal} {J. Phys. D: Appl. Phys.}\ }\textbf {\bibinfo
  {volume} {50}},\ \bibinfo {pages} {193001} (\bibinfo {year}
  {2017})}\BibitemShut {NoStop}%
\bibitem [{\citenamefont {Hohlfeld}\ \emph {et~al.}(2000)\citenamefont
  {Hohlfeld}, \citenamefont {Wellershoff}, \citenamefont {G{\"u}dde},
  \citenamefont {Conrad}, \citenamefont {J{\"a}hnke},\ and\ \citenamefont
  {Matthias}}]{Hohlfeld.2000}%
  \BibitemOpen
  \bibfield  {author} {\bibinfo {author} {\bibfnamefont {J.}~\bibnamefont
  {Hohlfeld}}, \bibinfo {author} {\bibfnamefont {S.-S.}\ \bibnamefont
  {Wellershoff}}, \bibinfo {author} {\bibfnamefont {J.}~\bibnamefont
  {G{\"u}dde}}, \bibinfo {author} {\bibfnamefont {U.}~\bibnamefont {Conrad}},
  \bibinfo {author} {\bibfnamefont {V.}~\bibnamefont {J{\"a}hnke}}, \ and\
  \bibinfo {author} {\bibfnamefont {E.}~\bibnamefont {Matthias}},\ }\bibfield
  {title} {\enquote {\bibinfo {title} {{Electron and lattice dynamics following
  optical excitation of metals}},}\ }\href@noop {} {\bibfield  {journal}
  {\bibinfo  {journal} {Chem. Phys.}\ }\textbf {\bibinfo {volume} {251}},\
  \bibinfo {pages} {237--258} (\bibinfo {year} {2000})}\BibitemShut {NoStop}%
\bibitem [{\citenamefont {Myroshnychenko}\ \emph {et~al.}(2008)\citenamefont
  {Myroshnychenko}, \citenamefont {Carb{\'o}-Argibay}, \citenamefont
  {Pastoriza-Santos}, \citenamefont {P{\'e}rez-Juste}, \citenamefont
  {Liz-Marz{\'a}n},\ and\ \citenamefont {Garc{\'\i}a~de
  Abajo}}]{Myroshnychenko.2008}%
  \BibitemOpen
  \bibfield  {author} {\bibinfo {author} {\bibfnamefont {V.}~\bibnamefont
  {Myroshnychenko}}, \bibinfo {author} {\bibfnamefont {E.}~\bibnamefont
  {Carb{\'o}-Argibay}}, \bibinfo {author} {\bibfnamefont {I.}~\bibnamefont
  {Pastoriza-Santos}}, \bibinfo {author} {\bibfnamefont {J.}~\bibnamefont
  {P{\'e}rez-Juste}}, \bibinfo {author} {\bibfnamefont {L.~M.}\ \bibnamefont
  {Liz-Marz{\'a}n}}, \ and\ \bibinfo {author} {\bibfnamefont {F.~J.}\
  \bibnamefont {Garc{\'\i}a~de Abajo}},\ }\bibfield  {title} {\enquote
  {\bibinfo {title} {{Modeling the Optical Response of Highly Faceted Metal
  Nanoparticles with a Fully 3D Boundary Element Method}},}\ }\href@noop {}
  {\bibfield  {journal} {\bibinfo  {journal} {Adv. Mater.}\ }\textbf {\bibinfo
  {volume} {20}},\ \bibinfo {pages} {4288--4293} (\bibinfo {year}
  {2008})}\BibitemShut {NoStop}%
\bibitem [{\citenamefont {Chen}\ \emph {et~al.}(2013)\citenamefont {Chen},
  \citenamefont {Shao}, \citenamefont {Li},\ and\ \citenamefont
  {Wang}}]{Huanjun.2013}%
  \BibitemOpen
  \bibfield  {author} {\bibinfo {author} {\bibfnamefont {H.}~\bibnamefont
  {Chen}}, \bibinfo {author} {\bibfnamefont {L.}~\bibnamefont {Shao}}, \bibinfo
  {author} {\bibfnamefont {Q.}~\bibnamefont {Li}}, \ and\ \bibinfo {author}
  {\bibfnamefont {J.}~\bibnamefont {Wang}},\ }\bibfield  {title} {\enquote
  {\bibinfo {title} {{Gold Nanorods and Their Plasmonic Properties}},}\ }\href
  {\doibase 10.1039/C2CS35367A} {\bibfield  {journal} {\bibinfo  {journal}
  {Chem. Soc. Rev.}\ }\textbf {\bibinfo {volume} {42}},\ \bibinfo {pages}
  {2679--2724} (\bibinfo {year} {2013})}\BibitemShut {NoStop}%
\bibitem [{\citenamefont {Balci}\ \emph {et~al.}(2014)\citenamefont {Balci},
  \citenamefont {Kocabas}, \citenamefont {K\"u\c{c}\"uk\"oz}, \citenamefont
  {Karatay}, \citenamefont {Akh\"useyin}, \citenamefont {Gul~Yaglioglu},\ and\
  \citenamefont {Elmali}}]{Balci.2014}%
  \BibitemOpen
  \bibfield  {author} {\bibinfo {author} {\bibfnamefont {S.}~\bibnamefont
  {Balci}}, \bibinfo {author} {\bibfnamefont {C.}~\bibnamefont {Kocabas}},
  \bibinfo {author} {\bibfnamefont {B.}~\bibnamefont {K\"u\c{c}\"uk\"oz}},
  \bibinfo {author} {\bibfnamefont {A.}~\bibnamefont {Karatay}}, \bibinfo
  {author} {\bibfnamefont {E.}~\bibnamefont {Akh\"useyin}}, \bibinfo {author}
  {\bibfnamefont {H.}~\bibnamefont {Gul~Yaglioglu}}, \ and\ \bibinfo {author}
  {\bibfnamefont {A.}~\bibnamefont {Elmali}},\ }\bibfield  {title} {\enquote
  {\bibinfo {title} {Probing ultrafast energy transfer between excitons and
  plasmons in the ultrastrong coupling regime},}\ }\href {\doibase
  10.1063/1.4892360} {\bibfield  {journal} {\bibinfo  {journal} {Appl. Phys.
  Lett.}\ }\textbf {\bibinfo {volume} {105}},\ \bibinfo {pages} {051105}
  (\bibinfo {year} {2014})}\BibitemShut {NoStop}%
\bibitem [{\citenamefont {Peruffo}, \citenamefont {Mancin},\ and\ \citenamefont
  {Collini}(2022)}]{Peruffo.2022}%
  \BibitemOpen
  \bibfield  {author} {\bibinfo {author} {\bibfnamefont {N.}~\bibnamefont
  {Peruffo}}, \bibinfo {author} {\bibfnamefont {F.}~\bibnamefont {Mancin}}, \
  and\ \bibinfo {author} {\bibfnamefont {E.}~\bibnamefont {Collini}},\
  }\bibfield  {title} {\enquote {\bibinfo {title} {{Ultrafast Dynamics of
  Multiple Plexcitons in Colloidal Nanomaterials: The Mediating Action of
  Plasmon Resonances and Dark States}},}\ }\href@noop {} {\bibfield  {journal}
  {\bibinfo  {journal} {J. Phys. Chem. Lett.}\ }\textbf {\bibinfo {volume}
  {13}},\ \bibinfo {pages} {6412--6419} (\bibinfo {year} {2022})}\BibitemShut
  {NoStop}%
\bibitem [{\citenamefont {Del~Fatti}, \citenamefont {Arbouet},\ and\
  \citenamefont {Vall{\'e}e}(2006)}]{delFatti.2006}%
  \BibitemOpen
  \bibfield  {author} {\bibinfo {author} {\bibfnamefont {N.}~\bibnamefont
  {Del~Fatti}}, \bibinfo {author} {\bibfnamefont {A.}~\bibnamefont {Arbouet}},
  \ and\ \bibinfo {author} {\bibfnamefont {F.}~\bibnamefont {Vall{\'e}e}},\
  }\bibfield  {title} {\enquote {\bibinfo {title} {{Femtosecond optical
  investigation of electron--lattice interactions in an ensemble and a single
  metal nanoparticle}},}\ }\href@noop {} {\bibfield  {journal} {\bibinfo
  {journal} {Appl. Phys. B: Lasers Opt.}\ }\textbf {\bibinfo {volume} {84}},\
  \bibinfo {pages} {175--181} (\bibinfo {year} {2006})}\BibitemShut {NoStop}%
\end{thebibliography}%

\end{document}